\documentclass[aps,pra,showpacs]{revtex4}

\usepackage[utf8x]{inputenc}
\usepackage{amsmath}
\usepackage{amssymb}
\usepackage{amsfonts}
\usepackage{graphics}
\usepackage[pdftex]{graphicx}
\usepackage[normal]{subfigure}
\usepackage{color}

\begin{document}

\title{Quantum Walks in artificial electric and gravitational Fields}
\author{Giuseppe Di Molfetta$\dagger$}
\author{M. Brachet*}
\author{Fabrice Debbasch$\dagger$} 
\affiliation{
$\dagger$ LERMA, UMR 8112, UPMC and Observatoire de Paris, 61 Avenue de l'Observatoire  75014 Paris, France\\
*Laboratoire de Physique Statistique, Ecole Normale Superieure, UPMC Universit\'e Paris 06, Universit\'e Paris Diderot, CNRS, 24 rue Lhomond, 75005 Paris, France}

\date{\today}

\begin{abstract}
The continuous limit of quantum walks (QWs) on the line is revisited through a recently developed method. In all cases but one, the limit coincides with the dynamics of a Dirac fermion coupled to an artificial electric and/or relativistic gravitational field. All results are carefully discussed and illustrated by numerical simulations. 
\end{abstract}

\maketitle
\section{Introduction}
QWs are simple formal analogues of classical random walks. They have been first considered by Feynman \cite{FeynHibbs65a} as possible discretizations of the free Dirac dynamics in flat space-time. They have been introduced in the physics literature by \cite{ADZ93a} and \cite{Meyer96a} and the continuous-time version first appeared in \cite{FG98a} . They have been realized experimentally in \cite{Schmitz09a, Zahring10a, Schreiber10a, Karski09a, Peruzzo10a, Sansoni11a, Sanders03a, Perets08a} and are important in many fields, ranging from fundamental quantum physics \cite{Perets08a, var96a} to quantum algorithmics \cite{Amb07a, MNRS07a}, solid state phsyics \cite{Aslangul05a, Bose03a, Burg06a, Bose07a} and biophysics \cite{Collini10a, Engel07a}.
Following Feynman's idea, several authors have studied the continuous limit of various QWs. The first publications \cite{BES07a, BH04a, Chandra10a, FeynHibbs65a, KRS03a, Strauch06a, Strauch07a, Strauch06b} only addressed QWs with constant coefficients and recent work has extended the discussion to QWs with time- and space-dependent coefficients \cite{DDMEF12a, DMD12a, DMD13a, DMD13b}, in both $(1 + 1)$ and $(1 + 2)$ space-time dimensions. In particular, a new method was developed in \cite{DDMEF12a, DMD12a, DMD13a, DMD13b} to investigate the continuous limit of QWs with non constant coefficients. This method delivers interesting results, not only for standard QWs, but also for `derived' QWs obtained from  original QWs by keeping only one time-step out of two \cite{DMD13b} . So far, this new method has only been applied to particular families of walks. This article presents the systematic application of this method to all QWs in $(1+1)$ space-time dimensions. The main conclusions are: (i) all families of walks do not admit a continuous limit (ii) when the limit exists, it coincides, in all cases but one, with the dynamics of a Dirac fermion coupled to an artificial electric field and/or relativistic gravitational field. These theoretical conclusions are illustrated by numerical simulations. Connections with previous results as well as other topics like transport in graphene are also discussed.

\section{Fundamentals}
\label{sec:Fund}

We consider quantum walks defined over discrete time and discrete one dimensional space, driven by
time- and space-dependent quantum coins acting on a two-dimensional Hilbert space $\mathcal H$. 
The walks are defined by the following finite difference equations, valid for all $(j, m) 
\in \mathbb{N}  \times \mathbb{Z}$: 
\begin{equation}
\begin{bmatrix} \psi^{L}_{j+1, m }\\ \psi^{R}_{j+1, m } \end{bmatrix} \  = 
B\left( \theta_{j, m} ,\xi_{j, m} ,\zeta_{j, m}, \alpha_{j,m} \right)
 \begin{bmatrix} \psi^{L}_{j, m+1} \\ \psi^{R}_{j, m-1} \end{bmatrix},
\label{eq:defwalkdiscr}
\end{equation}
where 
\begin{equation}
 B(\theta ,\xi ,\zeta, \alpha) = e^{i \alpha}
\begin{bmatrix}  e^{i\xi} \cos\theta &  e^{i\zeta} \sin\theta\\ - e^{-i\zeta} \sin\theta &  e^{-i\xi} \cos\theta
 \end{bmatrix}.
\label{eq:defB}
\end{equation} 
This operator is in $U(2)$, and is in $SU(2)$ only for $\alpha$ = $p \pi$, $p \in \mathbb {Z}$ and $\theta$, $\xi$ and $\zeta$ 
are then called the three Euler angles of $B$.
The index $j$ labels instants  and the index $m$ labels spatial points.
The two functions $\psi^L$ and $\psi^R$ can be interpreted as the components of a wave function $\Psi$ on a certain orthonormal basis $(b_L, b_R)$ independent of $j$ and $m$. 
These two components code for the probability amplitudes 
of the particle jumping towards the left or towards the right.
The total probability $\pi_j= \sum_m \left(
\mid \psi^L_{j, m} \mid ^2+ \mid \psi^R_{j, m}\mid^2 \right)$ is independent of $j$ {\sl i.e.} 
it is conserved by the walk.
The set of angles $\left\{ \theta_{j, m} ,\xi_{j, m} ,\zeta_{j, m},\alpha_{j,m} (j, m) \in \mathbb{N} 
 \times \mathbb{Z}\right\}$ 
defines the walks and is at this stage arbitrary.

Consider now, for all $(n, j) \in {\mathbb N}^2$, the collection 
$W_j^n = (\Psi_{k, m})_{k = nj, m \in \mathbb Z}$.
This collection represents the state of the quantum walk at `time' $k=n j$. For any given $n$, the collection $S^n = (W_j^n)_{j \in \mathbb Z}$ thus represents the entire history quantum walk observed through a stroboscope of `period' $n$. The evolution equations for $S^n$ are those linking $W_{j+1}^n$ to $W_j^n$ for all $j$. These can be deduced from the original evolution equations (\ref{eq:defwalkdiscr},\ref{eq:defB}) of the walk, which also coincide with the evolution equations of $S^1$. For example, the evolution equations of $S^2$ read:

\begin{eqnarray}
\psi^L_{j+2,m} & = & A^L_{j, m} \psi^L_{j,m+2} + B^L_{j, m} \psi ^L_{j,m} + C^L_{j, m} \psi^R_{j,m} + D^L_{j, m} \psi ^R_{j,m-2} \nonumber \\
\psi^R_{j+2,m} & = & A^R_{j, m} \psi^L_{j,m+2} + B^R_{j, m} \psi ^L_{j,m} + C^R_{j, m} \psi^R_{j,m} + D^R_{j, m} \psi ^R_{j,m-2} ,
\end{eqnarray}
where
\begin{eqnarray}
A^L_{j, m} & = & c_{j+1,m} \, c_{j,m+1} \  e^{i (\alpha_{j+1,m}+\xi_{j+1,m}+\alpha_{j,m+1} + \xi_{j,m+1})} \nonumber \\ 
B^L_{j, m} & = & -s_{j+1,m} \, s_{j,m-1} \  e^{i (\alpha_{j+1,m}+\zeta_{j+1,m}+\alpha_{j,m-1}- \zeta_{j,m-1})} \nonumber \\
C^L_{j, m} & = & c_{j+1,m} \, s_{j,m+1}  \  e^{i (\alpha_{j+1,m}+\xi_{j+1,m}+\alpha_{j,m+1} + \zeta_{j,m+1})} \nonumber \\
D^L_{j, m} & = & s_{j+1,m} \, c_{j,m-1} \  e^{i (\alpha_{j+1,m}+\zeta_{j+1,m}+\alpha_{j,m-1}- \xi_{j,m-1})} ,\nonumber \\
\end{eqnarray}
\begin{eqnarray}
A^R_{j, m} & = & -s_{j+1,m} \, c_{j,m+1} \   e^{i (\alpha_{j+1,m}-\zeta_{j+1,m}+\alpha_{j,m+1}+ \xi_{j,m+1})}  \nonumber \\ 
B^R_{j, m} & = &  - s_{j,m-1} \, c_{j+1,m}\   e^{i (\alpha_{j+1,m}-\xi_{j+1,m}+\alpha_{j,m-1}- \zeta_{j,m-1})} \nonumber \\
C^R_{j, m} & = & -s_{j+1,m}\, s_{j,m+1} \  e^{i (\alpha_{j+1,m}-\zeta_{j+1,m}+\alpha_{j,m+1} + \zeta_{j,m+1}}   \nonumber \\
D^R_{j, m} & = & c_{j,m-1}\, c_{j+1,m} \  e^{i (\alpha_{j+1,m}-\xi_{j+1,m}+\alpha_{j,m-1}- \xi_{j,m-1})},
 \nonumber \\
\end{eqnarray}
with $c_{jm} = \cos(\theta_{j,m})$ and  $s_{jm} = \sin(\theta_{j,m})$.

The QWs defined by (\ref{eq:defwalkdiscr}) admit a remarkable exact $U(1)$ gauge invariance. Consider indeed an arbitrary set of numbers $\{ \phi_{jm}, (j, m) \in \mathbb N \times \mathbb Z\}$, and write $\Psi_{jm} = {\Psi'}_{jm} \exp(i \phi_{jm})$. A straightforward computation shows that ${\Psi'}$ obeys
\begin{equation}
\begin{bmatrix} {\psi'}^{L}_{j+1, m }\\ {\psi'}^{R}_{j+1, m } \end{bmatrix} \  = 
B\left( {\theta'}_{j, m} ,{\xi'}_{j, m} ,{\zeta'}_{j, m}, {\alpha'}_{j,m} \right)
 \begin{bmatrix} {\psi'}^{L}_{j, m+1} \\ {\psi'}^{R}_{j, m-1} \end{bmatrix},
\label{eq:defwalkdiscrtilde}
\end{equation}
with
\begin{eqnarray}
{\alpha'}_{j, m} & = & \alpha_{j, m} + \frac{\sigma_{j,m}}{2}  \nonumber\\
{\xi'}_{j, m} & = & \xi_{j, m} + \delta_{j,m}  \nonumber\\
{\zeta'}_{j, m} & = & \zeta_{j, m} - \delta_{j,m}  \\
{\theta'}_{j, m} & = & \theta_{j, m} \nonumber
\label{eq:tildeangles}
\end{eqnarray}
and
\begin{eqnarray}
\sigma_{j,m} & = & \phi_{j, m+1} + \phi_{j, m-1} - 2 \phi_{j+1, m} \\
\delta_{j,m} & = & \frac{\phi_{j, m+1} - \phi_{j, m-1}}{2}.
\label{eq:sigmadelta}
\end{eqnarray}
It can also be shown that the $S^2$-type QW's admit the same discrete invariance.
As detailed below in Sections \ref{sec:n=1} and \ref{sec:n=2}, this discrete gauge invariance transcribes in the continuous limit into the standard continuous $U(1)$ gauge invariance of Maxwell electromagnetism.

To investigate the continuous limit of a collection $S^n$,
we first introduce a time step $\Delta t$ and a space step $\Delta x$.
We then consider that $\Psi_{jm}$ and $\theta_{jm}$ are the values taken 
by a two-component wave function $\Psi$ and by a function $\theta$ at the space-time point
$(t_j = j \Delta t, x_m = m \Delta x)$.
Thus, equation (\ref{eq:defwalkdiscr}) transcribes into:
\begin{equation}
\hspace{-0.5cm}
\begin{bmatrix} \psi^{L}(t_j+\Delta t, x_m) \\ \psi^{R}(t_j+\Delta t, x_m) \end{bmatrix} \  = 
B\left( \theta(t_j, x_m) ,\xi(t_j, x_m) ,\zeta(t_j, x_m), \alpha(t_j, x_m) \right)
 \begin{bmatrix} \psi^{L}(t_j, x_m+\Delta x) \\ \psi^{R}(t_j, x_m-\Delta x) \end{bmatrix}.
\label{eq:defwalk}
\end{equation}
We finally suppose, that $\Psi$ and $\theta$ are at least twice differentiable with respect to both space and time variables for all 
sufficiently small values of $\Delta t$ and $\Delta x$. 
The formal continuous limit of $S^n$ is defined as the couple of partial differential equations (PDEs) obtained from the discrete-time evolution equations defining $S^n$ 
by letting both $\Delta t$ and $\Delta x$ tend to zero.  

\section{How to determine the continuous limit}
\label{sec:Scaling}

Let us introduce a time-scale ${\mathcal T}$, a length-scale ${\mathcal L}$, an infinitesimal $\epsilon$ and write 
\begin{eqnarray}
\Delta t &=& \epsilon {\mathcal T} \nonumber \\ 
\Delta x &=& \epsilon^\delta {\mathcal L} \label{eq:scaling1},
\end{eqnarray}
where $\delta >0$ allows $\Delta t$ and $\Delta x$ to tend to zero differently.
We also allow the angles defining the walk to depend on $\epsilon$ and caracterize de $\epsilon$-dependance of these angles near $\epsilon = 0$ by the following
scaling laws:
\begin{eqnarray}
 \theta_\epsilon(t, x)& = \theta_0(t, x) + {\bar \theta}(t, x) \epsilon^\omega \nonumber  \label{eq:scaling2} \\
 \xi_\epsilon(t, x)& = \xi_0(t, x) + {\bar \xi}(t, x) \epsilon^\beta \nonumber \\
 \zeta_\epsilon(t, x)& = \zeta_0(t, x) + {\bar \zeta}(t, x) \epsilon^\gamma \\ 
 \alpha_\epsilon(t,x)& = \alpha_0(t, x) + {\bar \alpha}(t,x) \epsilon^\eta \nonumber 
\end{eqnarray}
where the four exponents $\omega$, $\beta$, $\gamma$ and $\eta$ are all positive. 
We also suppose that all functions are at least $C^2$ in $t$ and $x$.
The above relations define $1$-jets of quantum walks.  We finally denote by ${\mathcal B}_\epsilon (t, x)$ the matrix $B(\theta_\epsilon(t, x) ,\xi_\epsilon(t, x) ,\zeta_\epsilon(t, x), \alpha_\epsilon (t, x))$. 

Expand now the original discrete equations obeyed
by a jet $S^n_\epsilon$ around $\epsilon = 0$. A necessary and sufficient condition for the expansion to be self-consistent at order $0$ in $\epsilon$ is that
${\mathcal B}_0^n(t, x) = 1$ for all $t$ and $x$
(note from equation (1) that this condition is self-evident for $n = 1$). This transcribes into a constraint for the zeroth-order angles $\theta_0$, $\xi_0$, $\zeta_0$, $\alpha_0$.

Suppose this constraint is satified. The differential equations defining the continuous limit are obtained from the expansion by stating that the next lowest order contribution in $\epsilon$ identically vanishes. If one excepts zeroth-order terms, the terms of lowest order in the expansion scale as $\epsilon$, $\epsilon^\delta$, $\epsilon^\omega$, $\epsilon^\beta$, $\epsilon^\gamma$, $\epsilon^\eta$ (see for example the similar expansions performed on particular, simple quantums walks and presented in \cite{DMD12a, DMD13a} ). The richest and most interesting scaling is thus 
$\delta = \omega = \beta = \gamma = \eta = 1$, because this makes all the above terms of the same order and, thus, delivers a differential equation with a maximum number of contributions. This scaling will be retained in the remaindre of this article.

Note that  Equations (\ref{eq:scaling1}) and (\ref{eq:scaling2}) have actually very different meanings. Indeed, (\ref{eq:scaling1}) states that the relative variations of $\Psi$ between $j+1, m$ and $j, m \pm 1$ should be small, while (\ref{eq:scaling2}) states that the angles defining the walk do not deviate much from their zeroth-order values.

We will now present in detail the continuous limit of the jets $S_\epsilon^n$ for both $n = 1$ and $n = 2$. 

\section{Limit of $S^1_\epsilon$}
\label{sec:n=1}

\subsection{Zeroth order values of the angles}\label{sec:n=1_1}

The constraint on the zeroth-order angles reads:
\begin{eqnarray}
\sin \theta_0 & = & 0 \nonumber \\
e^{i (\alpha_0 + \xi_0)} \cos \theta_0 & = & 1 \nonumber \\
e^{i (\alpha_0 - \xi_0)} \cos \theta_0 & = & 1.
\end{eqnarray}
The above relations imply $\theta_0 = k \pi$, $\alpha_0 = (k + k_+ + k_-) \pi$, $\xi_0 = (k_+- k_-) \pi$, $(k, k_+, k_-) \in {\mathbb Z}^3$. The angle $\zeta_0$ does not enter this constraint and is therefore an arbitrary function of $t$ and $x$. For a given value of $\epsilon$, there is thus no meaningful distinction between $\zeta_0$ and $\zeta$. We will therefore from here on denote $\zeta_0$ by $\zeta$ in all equations, if only to simplify the notation.

\subsection{Equations of motion}\label{subss:eomel}

Let now $T = t/\mathcal{T}$, $X = x/\mathcal{L}$, $x^\pm = (T \pm X)/2$ and $\partial_\pm = \partial_{x^\pm} = \partial_T \pm \partial_X$. The variables $x^\pm$ are null coordinates in the  flat $2D$ space-time. With these notations, the equations of motion for the continuous limit of $S^1_\epsilon$ read:
\begin{eqnarray}
\partial_- \psi^L -  i( \bar \alpha + \bar \xi) \psi^L = + {\bar \theta} 
e^{i\left( \theta_0 + \alpha_0 + \zeta\right)} \psi^R \nonumber \\
\partial_+ \psi^R -  i( \bar \alpha - \bar \xi) \psi^R = - {\bar \theta} 
e^{i\left( \theta_0 + \alpha_0 - \zeta\right)} \psi^L,
\label{eq:Diracnequal1}
\end{eqnarray}
where $\theta_0$ and $\alpha_0$ are arbitrary multiples of $\pi$ (see the constraint above) and $\zeta$ is an arbitrary real function of $T$ and $X$.

Taken together, these two coupled first-order PDEs form  a Dirac equation in $(1 + 1)$ dimensions.
Let us indeed recall that, 
in flat two dimensional space-times, the Clifford algebra can be represented by $2 \times 2$ matrices 
acting on two-component spinors. This algebra admits two independents generators $\gamma^0$ 
and $\gamma^1$, which can be represented by $2 \times 2$ matrices obeying the usual anti-commutation relation:
\begin{equation}
\{\gamma^a, \gamma^b\} = 2 \eta^{ab} {\mathcal I},
\end{equation}
where $\eta$ is the Minkovski metric and $\mathcal I$ is the identity (unit) matrix. 
Consider the representation
$\gamma^0 = \sigma_1$ and $\gamma^1 = - \sigma_1 \sigma_3 = i \sigma_2$.
where $\sigma_1$, $\sigma_2$ and $\sigma_3$ are the three Pauli matrices:
\begin{equation}
\sigma_1 = 
\begin{bmatrix}  0 &  1\\ 1 &  0
\end{bmatrix}
,\ 
\sigma_2 = 
\begin{bmatrix}  0 &  -i\\ i &  0
 \end{bmatrix}
,\ 
\sigma_3 = 
\begin{bmatrix}  1 &  0\\ 0 &  -1
 \end{bmatrix}
\ .
\end{equation}
Equation (\ref{eq:Diracnequal1}) can be recast in the following compact form:
\begin{eqnarray}
 (i\gamma^0 D_0 + i \gamma^1 D_1 - \mathcal{M}) \Psi = 0
\label{eq:Diraccov}
\end{eqnarray}
where $D_\mu$ = $\partial_\mu$ - i $A_\mu$, $\partial_0 = \partial_T$, $\partial_1 = \partial_X$, 
$A_0 = \bar \alpha$, $A_1 = - \bar \xi$, 
${\mathcal M} = \mbox{diag}(m^-, m^+)$ and $m^\mp = \pm i\,  {\bar \theta} 
\exp
\left\{
i(\theta_0 + \alpha_0 \pm \zeta)
\right\}$. 

This equation describes the propagation in flat space-time of a Dirac spinor coupled to the Maxwell potential $A$ (the corresponding electric field is $E_X = \partial_T {\bar \xi} - \partial_X {\bar \alpha}$). The discrete gauge invariance presented in Section \ref{sec:Fund} degenerates accordingly into the standard local $U(1)$ invariance associated to electromagnetism. Indeed, suppose that the numbers $\phi_{jm}$ (see Section \ref{sec:Fund}) are the values taken by a function $\phi$ at space-time points $(t_j = j \Delta t, x_m = m \Delta x)$. Expanding equations (\ref{eq:tildeangles}) and (\ref{eq:sigmadelta}) at first order in $\epsilon$ delivers:
\begin{eqnarray}
{\alpha'} & = & \alpha - \epsilon\,  \frac{\partial \phi}{\partial T} 
\nonumber\\
{\xi'}  & = & \xi + \epsilon\,  \frac{\partial \phi}{\partial X}  \nonumber\\
{\zeta'} & = & \zeta - \epsilon\,  \frac{\partial \phi}{\partial X} \label{eq:tildeanglesc}\\
{\theta'} & = & \theta \nonumber.
\end{eqnarray}
The first two equations imply
\begin{eqnarray}
A'_0 = \frac{1}{\epsilon}\,\left(\alpha' - \alpha_0\right) = \frac{1}{\epsilon}\, 
\left(\alpha - \epsilon\,  \frac{\partial \phi}{\partial T} - \alpha_0\right) = 
A_0 - \frac{\partial \phi}{\partial T} \nonumber \\
A'_1 = \frac{1}{\epsilon}\,\left(\xi_0 - \xi'\right) = \frac{1}{\epsilon}\, 
\left(\xi_0 - \epsilon\,  \frac{\partial \phi}{\partial X} - \xi\right) = 
A_X - \frac{\partial \phi}{\partial X},
\label{eq:foo}
\end{eqnarray}
which are simply the standard gauge transformation for the potential $A$. The fourth relation implies that the mass tensor $\mathcal M$ is gauge invariant. Since the continuous limit equation of motion ({\ref{eq:Diracnequal1}) depends only on $\zeta_0$ (as opposed to $\zeta$), the third equation is not relevant to the continuous limit investigated in this Section. 

The angles $\theta_0$ and $\alpha_0$ are both multiples of $\pi$. Both masses are therefore complex conjugates to each other. They are real, and therefore identical, if $\zeta_0$ is an uneven multiple of $\pi/2$. They are both real and positive, equal to $\mid {\bar \theta} \mid$, if $\zeta_0 = \theta_0 + \alpha_0 + {\bar \sigma} + (2p + 1) \pi/2$, where $\exp(i {\bar \sigma}) = \mbox{sgn}\, {\bar \theta}$ is the sign of $\bar \theta$ and $p$ is an arbitrary integer. Note that, even in this case, the mass may depend on both $T$ and $X$.

\section{Limit of $S^2_\epsilon$}
\label{sec:n=2}

\subsection{Zeroth order values of the angles}

The constraint on the zeroth-order angles now reads:
\begin{eqnarray}
\cos \xi_0 \sin (2 \theta_0)  & = & 0 \nonumber \\
e^{2i \alpha_0} \left( e^{2i\xi_0} \cos^2 \theta_0 - \sin^2\theta_0 \right) & = & 1 \nonumber \\
e^{2i \alpha_0} \left( e^{- 2i\xi_0} \cos^2 \theta_0 - \sin^2\theta_0 \right) & = & 1.
\end{eqnarray}
As for $n = 1$, $\zeta_0$ does enter this constraint; it is therefore an arbitrary function of $t$ and $x$, which we denote simply by $\zeta$ (see the discussion at the end of Section \ref{sec:n=1_1}). 

The first relation implies that $\cos \xi_0 = 0$ (case 1) or $\sin 2 \theta_0 = 0$ (case 2). The first case corresponds to $\xi_0 = (2k + 1) \pi/2$, $k \in \mathbb Z$. The second and third relations then transcribe into the single constraint $\alpha_0 = (2k' + 1)\pi/2$, with $k' \in {\mathbb Z}$. Note that $\theta_0$ can then be an arbitray function of $t$ and $x$, as $\zeta_0$. This function will be simply denoted by $\theta$, just as $\zeta$ denotes $\zeta_0$.  

On the contrary, the second case corresponds to $\theta_0 = k \pi/2$, $k \in \mathbb Z$. 
If $k = 2p +1$, $p \in \mathbb Z$ (case 2.1), the last two constraint relations deliver simply $\alpha_0 = (2k' + 1)\pi/2$, with $k' \in {\mathbb Z}$. The angle $\xi_0$ is then arbitrary and will be denoted simply by $\xi$.
If $k = 2p$ (case 2.2), the last two constraint relations deliver $\alpha_0 = k' \pi/2$, $\xi_0 = \alpha_0 + k'' \pi$, $(k', k'') \in {\mathbb Z}^2$.

Cases 1, 2.1 and 2.2 partly overlap. Indeed, jets obeying $\theta_0 = k \pi$, $\xi_0 = (2k + 1) \pi/2$ and $\alpha_0 = (2k'+1) \pi/2$ can be filed under both case 1 and case 2. These are the only jets which can be filed under both cases.

Let us now give the equations of motion of the continuous limit in cases 1, 2.1 and 2.2.

\subsection{Equations of motion: case 1}
\label{ssec:n=2_case1}

\begin{eqnarray}
& & 2 \left( 
\partial_T - (\cos^2\theta) \partial_X \right) \psi^L  - 2i \left(({\bar \alpha} + (\cos^2 \theta) {\bar \xi} \right) \psi^L -  e^{+ i(\zeta - \xi_0)} (\sin 2 \theta)  \partial_X  \psi^R  =  
\nonumber \\
& & 
\left(- (\sin2 \theta) (\partial_X \theta) + i (\sin^2 \theta) (\partial_+ \zeta) \right) \psi^L 
+ \nonumber \\
& & + e^{+ i (\xi_0 + \zeta)} \left(
i (\partial_- \zeta)\frac{\sin 2 \theta}{2} - i {\bar \xi} (\sin 2 \theta) + (\partial_T \theta)
- (\partial_X \theta)(\cos 2 \theta)
\right) \psi^R
\end{eqnarray}
and
\begin{eqnarray}
& & 2 \left( 
\partial_T +  (\cos^2\theta) \partial_X \right) \psi^R  - 2i \left(({\bar \alpha} - (\cos^2 \theta) {\bar \xi} \right) \psi^R -  e^{- i(\zeta - \xi_0)} (\sin 2 \theta)  \partial_X  \psi^L  =  
\nonumber \\
& & 
\left(+ (\sin2 \theta) (\partial_X \theta) - i (\sin^2 \theta) (\partial_- \zeta) \right) \psi^R 
+ \nonumber \\
& & + e^{- i (\xi_0 + \zeta)}  \left(
 i (\partial_+ \zeta)\frac{\sin 2 \theta}{2} - i {\bar \xi} (\sin 2 \theta) - (\partial_T \theta)
- (\partial_X \theta)(\cos 2 \theta)
\right) \psi^L
\end{eqnarray}

These equations can be put into the more compact form
\begin{equation}
\partial_T \Psi + (\cos \theta) P \partial_X \Psi = Q \Psi,
\label{eq:DiracMR}
\end{equation}
where $\Psi = \psi^L b_L + \psi^R b_R$, 
\begin{equation}
P = \left(
\begin{matrix}
- \cos \theta & - e^{+i (\zeta - \xi_0)} \sin \theta \\ - e^{-i (\zeta - \xi_0)} \sin \theta & \cos \theta
\end{matrix}
\right)
\end{equation}
and
\begin{equation}
Q = \left(
\begin{matrix}
Q^L_L & Q^L_R \\
Q^R_L & Q^R_R
\end{matrix}
\right)
\end{equation}
with
\begin{eqnarray}
Q^L_L & = & i\left( {\bar \alpha} + (\cos^2 \theta) {\bar \xi} \right) - \frac{\sin 2\theta}{2} (\partial_X \theta) + \frac{i}{2} (\sin^2 \theta) (\partial_+ \zeta) \nonumber \\
Q^R_R & = & i\left( {\bar \alpha} - (\cos^2 \theta) {\bar \xi} \right) + \frac{\sin 2\theta}{2} (\partial_X \theta) - \frac{i}{2} (\sin^2 \theta) (\partial_- \zeta) \\
Q^L_R & = & \frac{e^{+ i (\xi_0 + \zeta)}}{2} \left(
i (\partial_- \zeta)\frac{\sin 2 \theta}{2} - i {\bar \xi} (\sin 2 \theta) + (\partial_T \theta)
- (\partial_X \theta)(\cos 2 \theta) \right) \nonumber \\
Q^R_L & = & \frac{e^{- i (\xi_0 + \zeta)}}{2}  \left(
 i (\partial_+ \zeta)\frac{\sin 2 \theta}{2} - i {\bar \xi} (\sin 2 \theta) - (\partial_T \theta)
- (\partial_X \theta)(\cos 2 \theta) \right) \nonumber
\end{eqnarray}

The operator $P$ is self-adjoint and its eigenvalues are $-1$ and $+1$. Two
eigenvectors associated to these eigenvalues are 
\begin{equation}
b_- = \left(\cos \frac{\theta}{2} \right) b_L + e^{i(-\zeta + \xi_0)}\left(\sin \frac{\theta}{2} \right) b_R,
\end{equation}
\begin{equation}
b_+ = \left(\sin \frac{\theta}{2}\right) e^{i(\zeta + \xi_0)} b_L + \left( \cos \frac{\theta}{2}\right) b_R.
\end{equation}
The family $(b_-, b_+)$
forms an orthonormal basis of the two dimensional spin Hilbert space, alternate to the original basis $(b_L, b_R)$. Let $\Psi = \psi^- b_- + \psi^+ b_+$. 
Equation (\ref{eq:DiracMR}) transcribes into:
\begin{eqnarray}
\partial_T \psi^- - (\cos \theta) \partial_X \psi^- - i \bar \alpha \psi^- - i \cos \theta \bar \xi \psi^- + \frac{i}{2}((\cos\theta-1)\partial_+\zeta)\psi^-  +
 \frac{\partial_X \theta}{2}\, (\sin \theta)\,  \psi^- & = & 0 \nonumber \\
\partial_T\psi^+ + (\cos \theta) \partial_X \psi^+ - i \bar \alpha \psi^+ + i \cos \theta \bar \xi \psi^+ - \frac{i}{2}((\cos\theta-1)\partial_-\zeta) \psi^+-\frac{\partial_X \theta}{2}\, (\sin \theta)\,  \psi^+ & = & 0.
\label{eq:definitive}
\end{eqnarray}

Suppose now, to make the discussion definite, that $\cos \theta$ is strictly positive and introduce in space-time $ \left\{(T, X) \right\}$ the Lorentzian, possibly curved metric $G$ defined by its covariant
components 
\begin{equation}
(G_{\mu \nu}) = \left(
\begin{matrix}
1 & 0\\0 & - \displaystyle{\frac{1}{\cos^2 \theta}}
\end{matrix}
\right),
\label{eq:metric}
\end{equation}
 where $(\mu, \nu) \in \left\{T, X \right\}^2$. 
This metric defines the canonical, scalar `volume' element 
${\mathcal D}_G X = \sqrt{- G}\,  dX = {dX}/{\cos \theta}$ in physical $1$D $X$-space, where $G$ is the determinant of the metric components $G_{\mu \nu}$.
Dirac spinors are normalized to unity with respect to ${\mathcal D}_G X$, whereas $\Psi$ is normalized to unity with respect to $dX$. We thus introduce $\Phi = \Psi \sqrt{\cos \theta}$ and rewrite the equations of motion (\ref{eq:definitive}) in terms of $\Phi$. We obtain:
\begin{eqnarray}
\gamma^a\left[ 
e^\mu_a D_\mu \Phi 
+ \, \frac{1}{2}\, 
\frac{1}{\sqrt{-G}}\,  \partial_\mu \left(
\sqrt{-G} e^\mu_a
\right) \Phi
\right] = 0,
\label{eq:DIRAC}
\end{eqnarray}
where $\mu \in \left\{T, X\right\}$, $a \in \left\{0, 1\right\}$ and $D_\mu = \partial_\mu - i A_\mu$ with 
\begin{equation}
A_T = {\bar \alpha} +  \frac{1 - \cos \theta}{2}\, \partial_X \zeta 
\label{eq:AT}
\end{equation}
and
\begin{equation}
A_X = - {\bar \xi} -  \frac{1 - \cos \theta}{2 \cos \theta}\, \partial_T \zeta.
\label{eq:AX}
\end{equation}
The usual $2D$ gamma matrices are:
\begin{equation}
\gamma^0= 
\left(
\begin{matrix}
0 & 1\\ 1 & 0
\end{matrix}
\right),\  
\gamma^1 = 
\left(
\begin{matrix}
0 & 1\\ - 1 & 0
\end{matrix}
\right),
\end{equation}
and the $e^\mu_a$ are the components of the diad (orthonormal basis)
$e_0 = e_T$ and $e_1 = \cos \theta\,  e_X$ on the original coordinate basis $(e_T, e_X)$. 
Equation (\ref{eq:DIRAC}) is the standard \cite{SR94a} equation of motion for a massless Dirac spinor propagating in $(1 + 1)$ dimensional space-time under the combined influence of the gravitational field $G$ and the electric field $E$ deriving from $A$.  Since the Dirac field is massless, its components are not coupled and evolve independently of each other. Each component follows a null geodesic of the gravitational field, and the electric field only modifies the energy along a given geodesic. Numerical simulations of a QW propagating radially in the gravitational field of a Schwarzschild black hole are presented below in Section \ref{sec:Schwarzschild}.

Let us conclude this section by commenting rapidly on how the discrete gauge invariance presented in Section \ref{sec:Fund} transcribes in the present context. The continuous limit equations (\ref{eq:tildeanglesc}) are of course valid. Combining these with (\ref{eq:AT}), (\ref{eq:AX}) and keeping only the lowest order terms in $\epsilon$ leads to the standard  gauge transformation $A'_0 = A_0 - \partial \phi/\partial T$ and $A'_1 = A_1 - \partial \phi/\partial X$. Just as it was the case in Section \ref{sec:n=1}, the transformation law for $\zeta$ does not contribute to the continuous gauge transformation, but it is not for the same reason. In Section \ref{sec:n=1}, the potential $A$ itself does not depend on $\zeta$. Here, the potential $A$ does depend on $\zeta$, but the gauge transformation for $\zeta$ generates terms of order $\epsilon$ in the gauge transformation for $A$, and these terms vanish as $\epsilon$ tends to zero. In the present context, the final, fourth equation in (\ref{eq:tildeanglesc}) reflects the fact  that the gravitational field does not depend on the choice of gauge for the phase of the spinor $\Psi$.

\subsection{Equations of motion: case 2.1}

The equations of motion of the continuous limit read:
\begin{eqnarray}
2\partial_T \psi^L - 2i {\bar \alpha} \psi^L & = & + i (\partial_+ \zeta) \psi^L +
2 {\bar \theta} e^{+ i \zeta} (\cos \xi)\, \psi^R
 \nonumber \\
 2\partial_T \psi^R - 2i {\bar \alpha} \psi^R & = & 
- i (\partial_- \zeta) \psi^R
- 2 {\bar \theta} e^{-i \zeta} (\cos \xi)\, \psi^L
\end{eqnarray}
where $\xi$ and $\zeta$ are arbitrary functions of $T$ and $X$. 
These equations are not PDEs, but {\sl ordinary} differential equations (ODEs) in $\psi^\pm$. Thus, there is for example no propagation in this case. Technically, this comes from the fact that $\theta_0$ is here constrained to be an uneven multiple of $\pi/2$.

\subsection{Equations of motion: case 2.2}

The equations of motion read:
\begin{eqnarray}
2 \partial_- \psi^L - 2 i \left({\bar \alpha} + {\bar \xi} \right) \psi^L 
& = & + 2 {\bar \theta} e^{i (2 \alpha_0 + \zeta)} (\cos \xi_0) \psi^R \nonumber \\
2 \partial_+ \psi^R - 2 i \left({\bar \alpha} - {\bar \xi}\right) \psi^R 
& = & - 2 {\bar \theta} e^{i (2 \alpha_0 - \zeta)} (\cos \xi_0) \psi^L
\label{eq:Diracnequal2}
\end{eqnarray}
where $\alpha_0$ is a multiple of $\pi/2$, $\xi_0 - \alpha_0$ is multiple of $\pi$ and $\zeta$ is an arbitrary function of $T$ and $X$.
Equation (\ref{eq:Diracnequal2}) can be recast in the following compact form:
\begin{eqnarray}
 (i\gamma^0 D_0 + i \gamma^1 D_1 - \mathcal{M}) \Psi = 0
\end{eqnarray}
where $D_\mu$ = $\partial_\mu$ - i $A_\mu$, $\partial_0 = \partial_T$, $\partial_1 = \partial_X$, 
$A_0 = \bar \alpha$, $A_1 = - \bar \xi$, 
${\mathcal M} = \mbox{diag}(m^-, m^+)$ and $m^\mp = \pm i\,  {\bar \theta} 
\exp
\left\{
i(2\alpha_0 \pm \zeta)
\right\} (\cos \xi_0)$. 
This equation describes the propagation in flat space-time of a Dirac spinor $\Psi$ coupled to the potential $A$ and with a mass tensor $\mathcal M$.

\section{Numerical simulations}

\subsection{Basics}\label{ssec:Bas}

In order to ascertain the validity of the continuous limits that were derived above, we wish to compare numerical solutions of the QW defined by the finite difference equations (\ref{eq:defwalkdiscr},\ref{eq:defB}) with the  corresponding Dirac-type PDEs defined by  Eqs.(\ref{eq:DIRAC})  and (\ref{eq:Diraccov}).

While the numerical integration of the QW finite difference equations poses no particular problem, controlling the error on numerical solutions of PDEs is a more involved matter. This hurdle can be avoided in the special case where the mass terms cancels, because one can then compare the numerical solutions of the QW  finite difference equations with the numerical solutions of the ODEs defining the \emph{characteristics} of the masless Dirac PDE (see below sec.\ref{sec:Schwarzschild}).

We have chosen to use Fourier pseudo-spectral methods \cite{Got-Ors}, for their precision and ease of implementation.
We therefore restrict ourself to $2 \pi$-periodic boundary conditions.
A generic field $\psi(x)$ is thus evaluated on the $n$ collocation points $x_j=2 \pi j/n$, with $j=0,n-1$ as  $\psi_j=\psi(x_j)$.
The discrete Fourier transforms are standardly defined as 
\begin{eqnarray}
\psi(x_j)&=&\sum_{k=-n/2}^{n/2-1} \exp{(i k x_j)}\hat{\psi}_k \nonumber \\
\hat{\psi}_k&=& \frac{1}{n} \sum_{j=0}^{n-1} \psi(x_j) \exp{(-i k x_j)}
 \label{eq:DFT}
\end{eqnarray}
These sums can be evaluated in only $n \log(n)$ operations by using Fast Fourier Transforms (FFTs).
Spatial derivatives of fields are evaluated in spectral space by multiplying by $i k$ and products  are evaluated in physical space.

The original QW equations (\ref{eq:defwalkdiscr},\ref{eq:defB})  can also be simply cast in this setting, as the translation operator 
$\psi_j \to \psi_{j \pm 1}$ is represented in Fourier space by $\hat{\psi}_k \to \hat{\psi}_k  \exp{(\pm i k 2 \pi /n)}$. 
In this setting, the continuous limit is automatically taken when $n$ is increased.

\subsection{QWs in constant a uniform electric field}

\begin{figure}[ht!]
\subfigure{\includegraphics[scale=0.4]{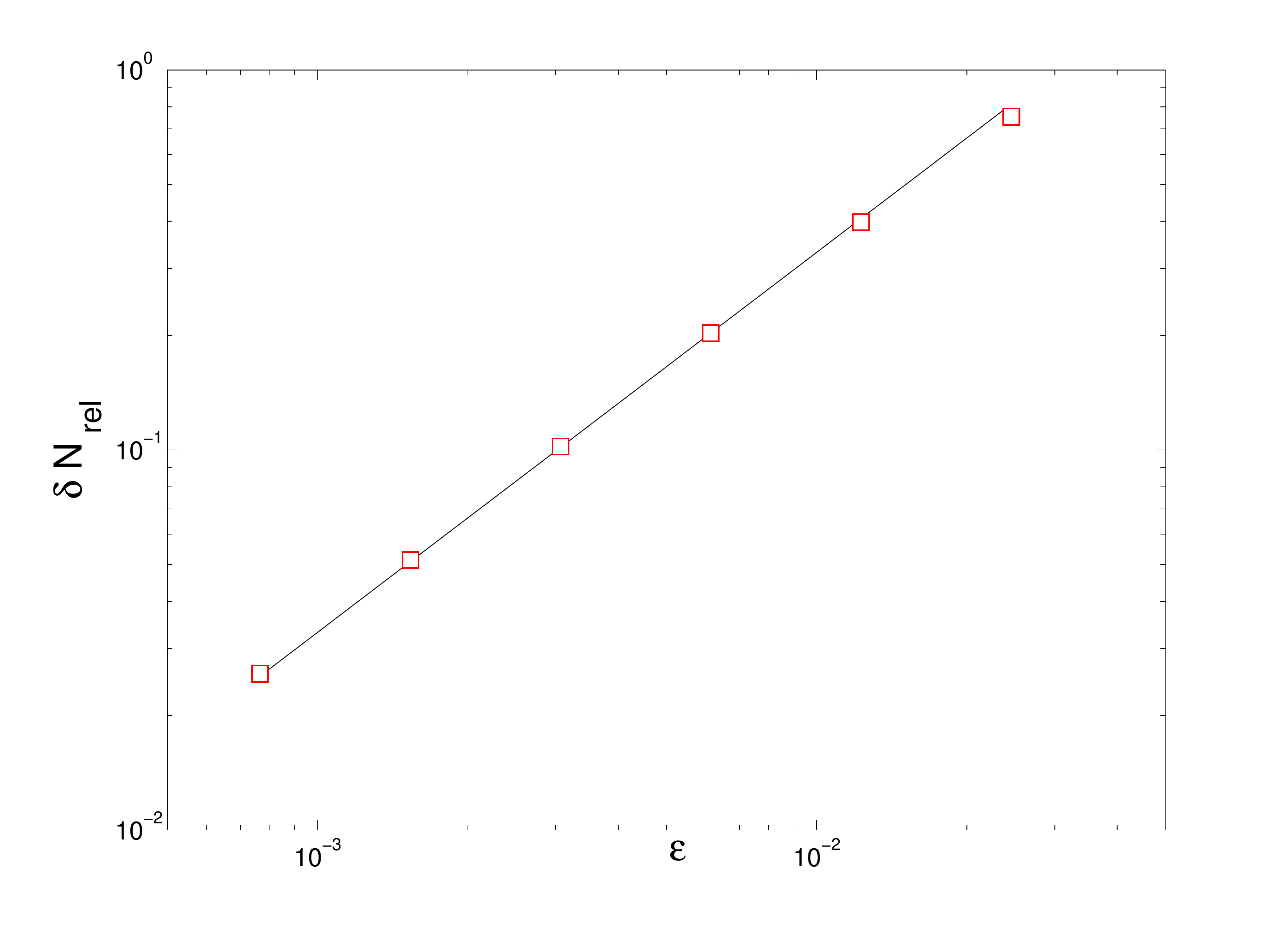}}
\caption{Relative difference between the density of the  
QW and the density of the solution of the Dirac equation $\delta N_{\rm rel}=\sqrt{<(N_{QW}-N_{D})^2>}/<(N_{D})>$ plotted at T = 100 versus $\epsilon=\Delta X$ in Log-Log representation. We have plotted the relative difference in the same conditions as in Fig.\ref{Fig:density}.a below but for five different resolutions (i.e. value of $\epsilon$): from the left n = $2^8$, $2^9$,$2^{10}$, $2^{11}$, $2^{12}$, $2^{13}$. The solid black line represents the expected $\epsilon^1$ scaling.
}\label{Fig:error}
\end{figure}

As explained in Sections II, the QWs and the Dirac equation exhibit a $U(1)$ gauge invariance. All choices of gauge thus correspond to the same physics. Within a pseudospectral code, the right gauge to work with a $1D$ constant uniform electric field $E$ is $A_0= 0$, $A_1 = -E \ T$; in particular, the other `natural choice' $A_0= -E \ X$, $A_1$ breaks the spatial periodicity condition. In all QW simulations, the retained choice of gauge has been implemented by choosing the following numerical values:

\begin{eqnarray}
\alpha(T, X) & = & 0  \nonumber \\
\xi(T, X) & = & 1.1 \ T \nonumber\\
\zeta (T, X) & = & \frac{\pi}{2} \nonumber \\
\theta(T, X) & = & 0.24 \nonumber.
\label{eq:anglesnumel}
\end{eqnarray}

\begin{figure}[ht!]
\subfigure{\includegraphics[scale=0.5]{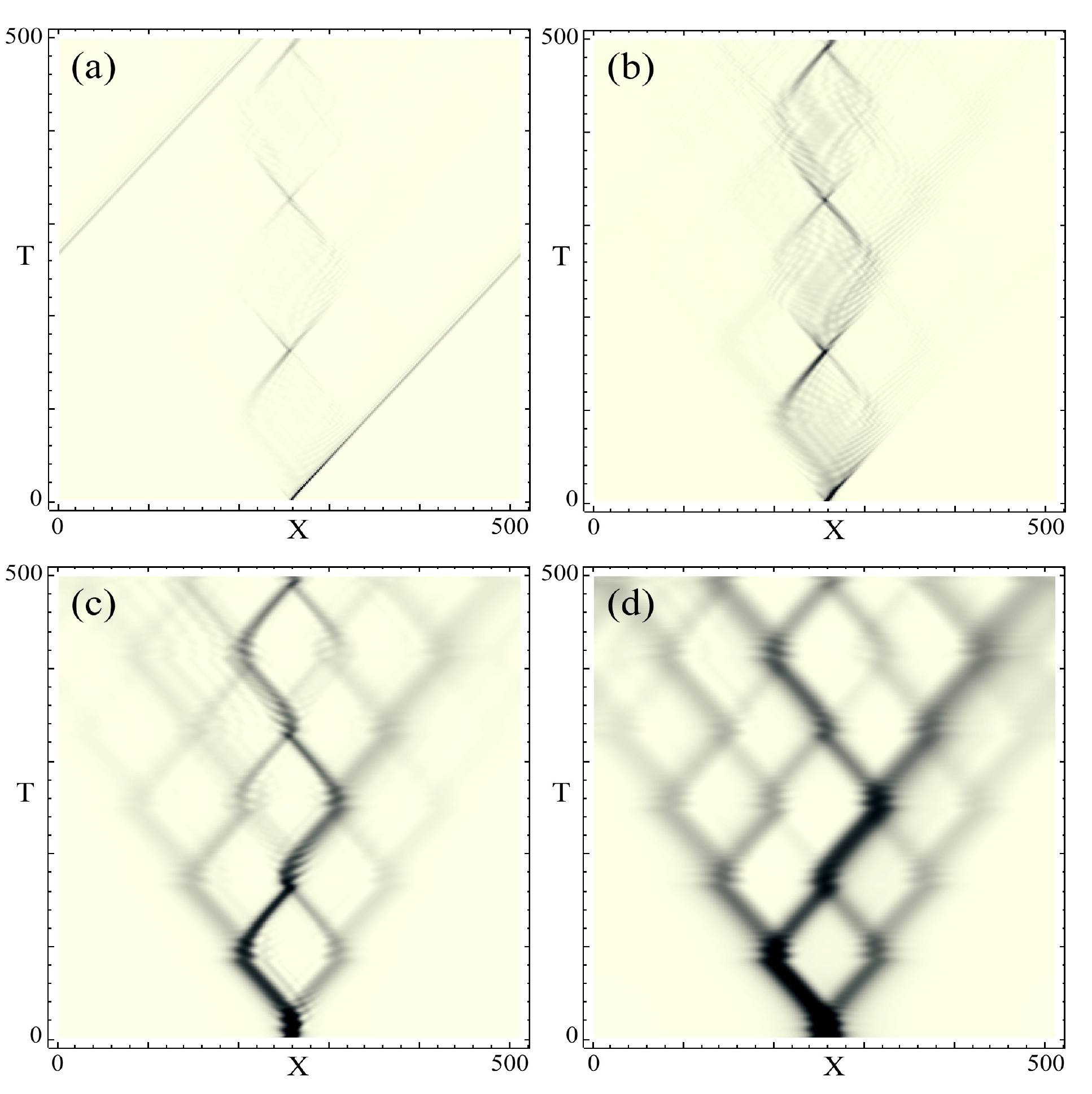}}
\caption{Quantum simulation of equations (\ref{eq:defwalkdiscr},\ref{eq:defB}) representing a Dirac particle in a constant and static electric background (see section \ref{subss:eomel} and Eq.(\ref{eq:anglesnumel})). The initial condition is a Gaussian wave packet of positive energy solutions with width $\sigma_X$ = 0.005 in (a), 0.01 in (b), 0.03 in (c), 0.08 in (d) and resolution $n=2^9$.}\label{Fig:density}
\end{figure}

We used initial data consisting in gaussian wave packets of positive energy solutions to the free Dirac equation. The gaussian widths $\sigma_X$ are such that they are well resolved within the used resolutions so that spectral convergence is ensured.

As discussed above (see end of section \ref{ssec:Bas}) the QW and its Dirac continuous limit can be jointly simulated within the same pseudo spectral algorithm. This allows for a very simple, direct evaluation of the discrepancy between the QW and the corresponding solution of the Dirac equation. This discrepancy can be measured by the relative difference $\delta N_{\rm rel}$ between the density of the  
QW and the density of the solution of the Dirac equation.

Figure \ref{Fig:error} shows that such a typical relative difference scales as $\epsilon$, as expected. Indeed,  for a single time-step, the discrepancy is theoretically of order $\epsilon^2$. Thus, after a {\sl fixed time $T = O(\epsilon^{-1})$}, the discrepancy is of order $\epsilon^{-1} \epsilon^2 = \epsilon$. 

This result confirms that QWs can be used to simulate Dirac dynamics in constant electric field, as was done for examaple in \cite{witt10a, Longhi10a}. Both QW and Dirac dynamics are very rich, as exemplified by Figure \ref{Fig:density}, which compare with Fig. 2 of ref.\cite{witt10a} and Fig. 3 of ref.\cite{Longhi10a}. 

Note that, as $\sigma_X$ increases, the spatial dispersion of the
wave packet also increases which makes the time evolution of the density more complex. The solution which is initially a positive energy planar wave start to oscillate between positive and negative modes under the action of the constant electric field displaying high-frequency Zitterbewegung in Fig.\ref{Fig:density}.c-d. 
Offering new results of the Dirac dynamics in presence of an electric field is not the purpose of this article. Let us conclude this Section by offering instead a brief historical overview of the very large litterature already existing on the topic.

In 1929, Klein studied a relativistic scalar particle moving in an external step function potential. He found a paradox that, in the case of a strong potential, the reflected flux is larger than the incident flux although the total flux is conserved \cite{Klein1929}.
Sauter studied this problem for a Dirac spin $1/2$ particle by considering a potential corresponding to a electric field with constant value $E_0$ on a given interval. He found an expression for the transmission coeffcient of the wave through the electric potential barrier from the negative energy state to positive energy states \cite{Sauter1931}.
This remarkable phenomenon was related, in 1936, to positron-electron pair creations by Heisenberg and his student Hans Euler \cite{HeisenbergEuler1936}.

Of course, in order to deal with anti-particles a massive reinterpretation of the Dirac equation theory is necessary \cite{BjorkenDrell}, leading to modern field theory and quantum electrodynamics.
The modern formula for pair creation in a constant external electric field was delivered, in 1951, by Schwinger \cite{Schwinger1951}. It involves the same dominant exponential term $\exp(-\pi \frac{m_e^2c^3}{\hbar e E})$ that was derived, 20 years before, by Sauter. 
A detailed review of these historical developments  is given in the first sections of reference \cite{Ruffini2010}.

\subsection{QWs in Schwarzschild black hole}
\label{sec:Schwarzschild}
A Schwarschild black hole is a spherically symmetric solution of Einstein equation in vacuo. 
The corresponding $4D$ metric reads, in dimensionless Lema\^itre coordinates $(\tau, \rho, \theta, \phi)$ \cite{L33a}:
\begin{equation}
ds^2 = d\tau^2- \frac{r_g}{r}\,  d\rho^2 - r^2 d\Omega
\end{equation}
where 
$r (\tau, \rho) = r_g^{1/3} \left[ \frac{3}{2}\left(\rho - \tau\right)\right]^{2/3}$,
$d\Omega = d\theta^2 + (\sin^2 \theta) d\phi^2$. The event horizon is located at $r = r_g$ {\sl i.e.} $\rho = \tau + (2/3) r_g$, and the singularity is located at $r = 0$ {\sl i.e.} $\rho = \tau$. The exterior of the black hole is the domain $r > r_g$.The range of variations for the Lema\^itre coordinates is $\tau \ge 0$, $\rho \ge \tau$ ({\sl i.e.} $r (\tau, \rho) \ge 0$), $0 \le \theta \le \pi$, $0 \le \phi < 2 \pi$.

Because of the spherical symmetry, a point mass which starts its motion radially will go on moving radially. Radial motion can be studied by introducing the $2D$ metric $g$, also singular at $r=0$, with covariant components $g_{\tau \tau} = 1$, $g_{\rho \rho} = - r_g/r$, $g_{\tau \rho} = g_{\rho \tau} = 0$.
The null geodesics of $g$ are defined by $d \tau = \pm \left( r_g/r(\tau, \rho)\right)^{1/2} d\rho$. Note that the $2D$ projection of the horizon on the $(\tau, \rho)$-plane coincides with a null geodesics of $g$.

We now identify the dimensionless time $T$ with the time coordinate $\tau$ and the dimensionless space variable $X$ with $\lambda \rho$, where 
$\lambda$ is an arbitrary strictly positive real number (see Fig.\ref{Fig:QW}). The `radius' $r$ can then be expressed as a function of $T$ and $X$:
\begin{equation}
r (T, X) = \left[\frac{3}{2}\left(\frac{X}{\lambda} - T\right)\right]^{2/3}r_g^{1/3},
\label{eq:defr}
\end{equation}
and the components of $g$ in the coordinate basis associated to $T$ and $X$ are
$g_{TT} = 1$, $g_{XX} = - r_g/(\lambda^2 r)$, $g_{TX} = g_{XT} = 0$. 
Note that the condition $\rho \ge \tau$ transcribes into $X \ge \lambda T$. 

Let $\mathcal D$ be the domain where $-g_{XX} \ge1$. This domain is characterized, in $(T, X)$ coordinates, by the condition
\begin{equation}
X \leq \lambda T + \frac{2}{3 \lambda^2}\, {r_g}.
\end{equation}
In $\mathcal D$ the metric $g$ can be identified with the metric $G$ (see Eq.(\ref{eq:metric})).
This identification defines an angle $\theta$ which depends on $T$ and $X$ by: 
\begin{equation}
(\cos \theta) (T, X) = \lambda\, \sqrt{\frac{r(T, X)}{r_g}}
\label{eq:deftheta}
\end{equation}.
A QW in $\mathcal D$ can de defined by complementing this choice of $theta$ by a choice of the other three angles. All simulations were done with

\begin{eqnarray}
\alpha(T, X) & = & 0  \nonumber \\
\xi(T, X) & = & \pi \nonumber\\
\zeta (T, X) & = & \frac{\pi}{2} \nonumber \\
\label{eq:anglesnum}
\end{eqnarray}

This QW has already been considered in \cite{DMD13b}.

The condition defining $\mathcal D$ can be rewritten as $r \le r_g/\lambda^2$, The domain $\mathcal D$ thus includes, for all $\lambda$, the singularity located at $r = 0$. For $\lambda >1$, $r_g/\lambda^2 < r_g$ and $\mathcal D$ is then entirely located inside the horizon.
For $\lambda = 1$, $\mathcal D$ coincides with the interior of the horizon, and $\mathcal D$ extends outside the horizon for $\lambda < 1$. 

Ref. \cite{DMD13b} offers plots of she spatial density $\mid \Psi(T, X) \mid^2$ for several initial conditions. These plots confirm that the QW follows to a great accuracy the radial null geodesics of the Schwarzschild metric, except perhaps as the QW approaches the singularity. 
This phenomenon is explored in detail by Figure 3. The plots reveal the existence of interesting `interferences' near the singularity (see (a1) and (a2)), which seem to disappear as $\epsilon$ tends to zero.

\begin{figure}[ht!]
\subfigure{\includegraphics[scale=0.5]{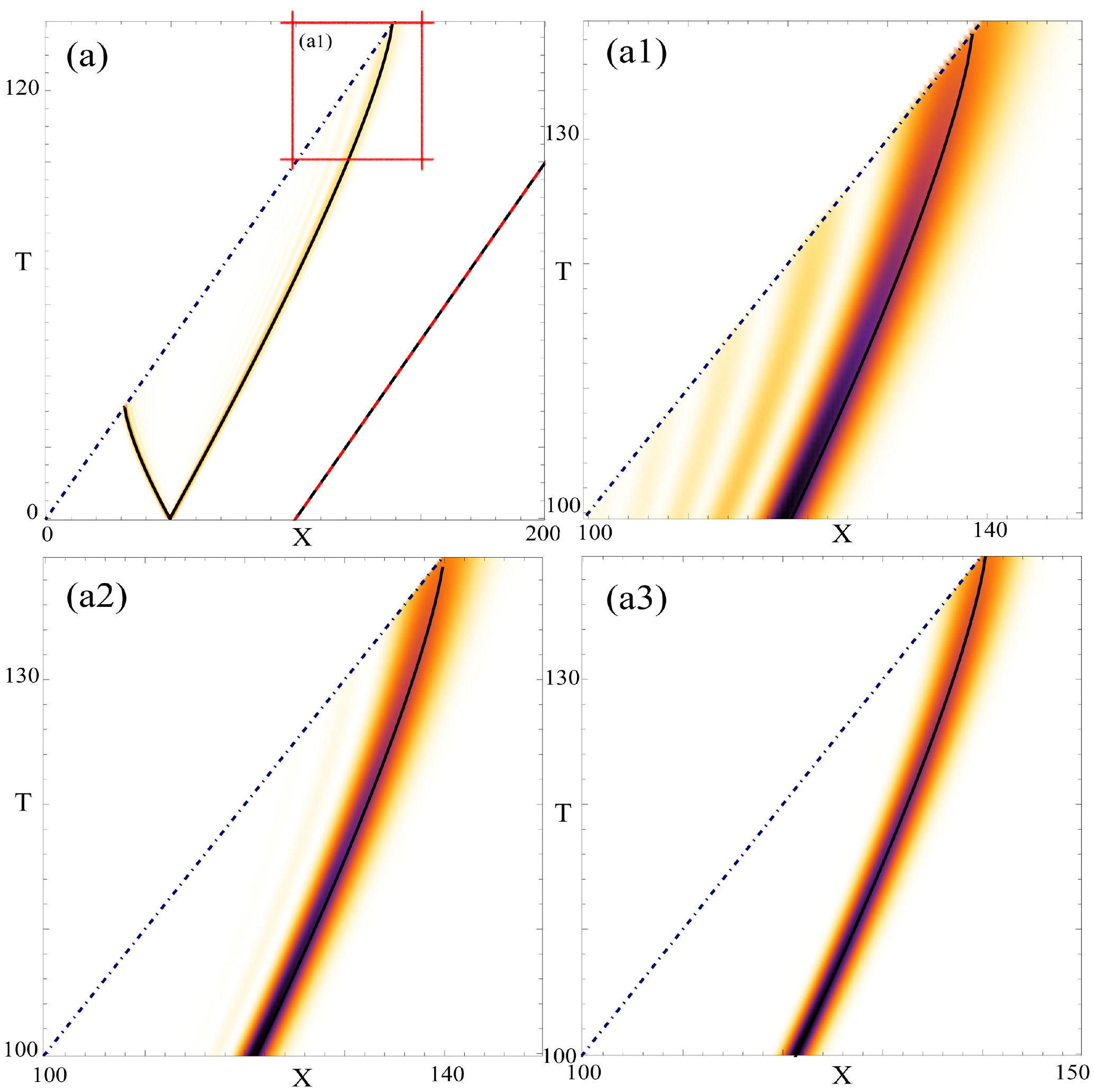}}
\caption{(color online)Time evolution density of the QW vs. null geodesics (solid curves) of the 2D Schwarshild metric with $\lambda = 1$. 
The initial condition is $\Psi(0,X) = \sqrt{N_0}(X) (b_L + I b_R)$ with an initial Gaussian density $N_0$ of $\sigma_X = 0.5$ centered on $X_0= 50.5$.
The singularity is represented by the dotted and dashed line on the left and the horizon (which is a null geodesique) is represented by the dashed line. The two branches of the QW which starts inside the horizon end up on the singularity. The (red) solid line represents the limit of the definition domain $\mathcal{D}$ of the QW. In (a1) note that the right branch of the QW lags slightly behind the null geodesic when approaching the r=0 singularity. The agreement between the geodesics and the density profile of the walk gets better as we increase the resolution of the simulation: $200$ gridpoints in (a1), $800$ in (a2) and $1600$ in (a3).}
\label{Fig:QW}
\end{figure}

\section{Conclusion}

We have revisited the continuous limit of discrete time QWs on the line, keeping every step or only one step out of two. We have identified all families of walks which admit a continuous limit and obtained the associated PDEs. In all cases but one, the PDE describes the propagation of a Dirac fermion coupled to an electric field and, possibly, to a general relativistic gravitational field. We have also illustrated these conclusions by new numerical simulations.

Let us now discuss rapidly the above results. 

As mentionned in the introduction, all above literal computations are based on a new method first introduced in \cite{DMD12a, DMD13a, DMD13b, DDMEF12a}. New to this article is all the material presented in Section \ref{sec:n=2}. The Dirac equation obtained in Section \ref{sec:n=1} has already been presented in \cite{DMD12a, DMD13a, DMD13b, DDMEF12a} , but without the important discussion of the $U(1)$ gauge invariance. The discrete gauge invariance presented in Section \ref{sec:Fund} is also new. Let us mention in this context that QWs coupled to a uniform and constant electric field have also been considered in \cite{mesch13a} . These so-called `electric walks' are particular cases of the walks considered in \cite{DMD12a, DMD13a, DMD13b, DDMEF12a}  and in Section \ref{sec:n=1} of the present article. In \cite{mesch13a}, the constant and uniform electric field is put by hand on the equations of motion of the walks. On the contrary, the approach developed in the present article makes it clear that the electric field is simply a manifestation of the time-and space-dependance of the angles defining the walks. This approach also allows for a straightforward generalization to non constant and/or nonuniform electric fields (Sections \ref{sec:n=1} and \ref{sec:n=2}), and to gravitational fields (\ref{sec:n=2}). The electric and gravitational fields coupled top the QWs thus clearly appear as synthetic gauge fields \cite{dalibard10a}  .

The work presented in this article should be extended in several directions. One should first determine how the new method works, and what it delivers, when one keeps only one step out of $n$ for arbitrary $n >2$. Extensions to higher dimensional space and/or to higher dimensional Hilbert space are also desirable. In particular, the fact that some QWs on the line can be interpreted as the propagation of charged massless Dirac fermions suggests that QWs could be useful in modeling charge transport in graphene \cite{Novo05, elia11}. Let us note that the irinherent discreteness would give QWs a strong computational advantage over the more traditional models based on PDEs. Finally, determining systematically the continuus limit of non linear QWs \cite{perez10} and of walks in random media \cite{chandra13} should also prove interesting.

\end{document}